\documentclass{ws-procs975x65}            % default, citations in superscript
\newcommand{\be}{\begin{equation}}
\newcommand{\ee}{\end{equation}}

\newcommand{\rL}{\rho_\Lambda}

\newcommand{\CC}{\Lambda}

\newcommand{\rLo}{\rho_{\CC}^0}

\begin{document}
\title{Signs of Interacting Vacuum and Dark Matter in the Universe}

\author{Joan Sol\`a Peracaula\footnote{Invited talk at MG 15, U.  Rome  ``La Sapienza''. Updated presentation.}, Adri\`a G\'omez-Valent and Javier de Cruz P\' erez}

\address{%High Energy Physics Group,
Departament de F\'isica Qu\`antica i Astrof\'isica (FQA), \\  and Institute of Cosmos Sciences (ICCUB)\\ Universitat de Barcelona, Av. Diagonal 647, E-08028 Barcelona, Catalonia, Spain \\
sola@fqa.ub.edu, adriagova@fqa.ub.edu, decruz@fqa.ub.edu}

\begin{abstract}
We consider the impact of dynamical dark energy (DDE) in the possible solution of the existing tensions in the $\Lambda$CDM. We test both interacting and non-interacting DE models with dark matter (DM). Among the former, the running vacuum model (RVM) interacting with DM appears as a favored option.  The non-interacting scalar field model  based on the  potential $V\sim \phi^{-\alpha}$,  and the generic XCDM parametrization, also provide consistent signs of DDE. The important novelty of our analysis with respect to the existing ones in the literature is  that we use the matter bispectrum, together with the power spectrum. Using a complete and updated set of cosmological observations on SNIa+BAO+$H(z)$+LSS+CMB,  we find that the crucial triad BAO+LSS+CMB  (i.e.  baryonic acoustic oscillations, large scale structure formation data and the cosmic microwave background)  provide the bulk of the signal. The bispectrum data is  instrumental to get hold of the DDE signal, as our analysis shows. If the bispectrum is not included, the DDE signal could not be currently perceived at a significant confidence level.
\end{abstract}


\bodymatter

\section{Introduction}\label{introduction1}

The cosmological constant problem in the context of quantum field theory (QFT) is well known. Already in the standard model of particle physics the value of the vacuum energy density $\rL$ is predicted to be some $55$ orders of magnitude higher than the measured value, $ \rL\sim 10^{-47}$ GeV$^4$, see \cite{CCPWeinberg89,PeeblesRatra2003,JSPReview2013,SolGomRev2015,AdriaPhD2017}.
Rather than attempting to solve this problem here, we adopt a pragmatic point of view and consider the possibility to test whether the vacuum energy, and in general the dark energy (DE), is dynamical or not in the light of the current observations. This might help to shed some light not only on how to cure the CC-problem eventually, but also on how to solve some of the existing tensions between the $\CC$CDM model and observations, like the $H_0$-tension (the mismatch between the Planck measurements of $H_0$ and the local measurements based on the distance ladder approach)\cite{Planck2015,Planck2018,RiessH0201618} and the $\sigma_8$-tension, which is described in terms of the combined observable $f(z)\sigma_8(z)$, where $f(z)$  is the linear growth rate and $\sigma_8(z)$ is  the RMS matter fluctuation on scales of $R_8=8 h^{-1}$ Mpc at redshift $z$. The corresponding prediction of the $\CC$CDM is known to be exceedingly large and hence the description of the LSS data  is poorly accounted for by the concordance model \cite{Macaulay2013,BasilakosNesseris2017}. In the following we briefly review how dynamical dark energy (DDE) can lend a helping hand on trying to solve some of these problems, although the evidence of DDE (or lack of it) varies significantly if taking different studies in the literature, see e.g. \cite{ApJMPLA,PLB2017,MNRAS2018a,MNRAS-EPL2018,GBZhao2017,CostaLi,ParkRatraOoba} and the more recent analyses\,\cite{PDU2019,Martinelli2019}, and references therein.

\section{Cosmology beyond $\CC=$const.}
The Friedmann and acceleration equations when the vacuum evolves with the cosmic expansion, $\rL=\rL(t)$ can be written formally similar to the standard case when $\rL=$const. If we focus on flat FLRW metric, the field equations that incorporate vacuum dynamics read
 \begin{equation}\label{eq:generalizedFriedmann}
3H^2=8\pi
G(\rho_m+\rL(t))\,,\ \ \ \ \ \ \ \  2\dot{H}+3H^2=-8\pi
G(w_m\rho_m+w_\CC\rL(t))\,,
\end{equation}
where  $w_m=1/3,0$ for relativistic and nonrelativistic matter, and  $w_\CC=-1$ for the vacuum (irrespective of its time evolution).  Hereafter we assume $w_\CC=-1$ -- see \cite{PLB2017} for the analysis of the general case with  $w_\CC=-1+\epsilon$ ($|\epsilon|\ll 1$), i.e. for quasivacuum models. The total matter density $\rho_m$ involves the contributions from baryons and cold dark matter (DM): $\rho_m=\rho_b+\rho_{dm}$. We assume that the DM component is the only one that interacts with vacuum, whilst radiation and baryons are taken to be self-conserved, hence $\rho_r(a)=\rho_{r}^0\,a^{-4}$ and $\rho_b(a) = \rho_{b}^0\,a^{-3}$, where the superscript zero denotes the current values.  The DM component, instead, exchanges energy with the vacuum, and the local conservation law reads
\begin{figure}[t]
\centering
\includegraphics[height=4.5cm]{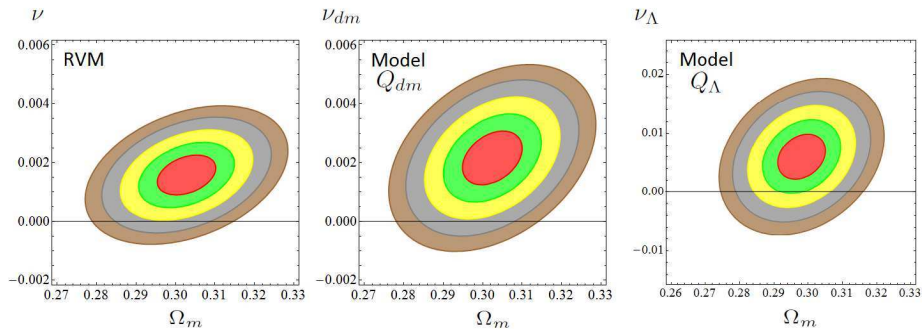}
\caption{Likelihood contours for the three DVMs (3)-(5) in the $(\Omega_m,\nu_i)$ plane up to 5$\sigma$ c.l. after marginalizing over the rest of the fitting parameters. Use is made of SNIa+BAO+$H(z)$+LSS data and compressed CMB  Planck 2015 data. The $\CC$CDM ($\nu_i=0$) appears  disfavored  in front of the DVMs. In particular, the RVM and the $Q_{dm}$ are favored at $\sim 3\sigma$ c.l. See\,\cite{MNRAS-EPL2018} for details on the SNIa+BAO+$H(z)$+LSS+CMB data used in our fit. We note, in particular, that bispectrum data is also included\,\cite{GilMarin2016}.}
\end{figure}
\begin{equation}\label{eq:Qequations}
\dot{\rho}_{dm}+3H\rho_{dm}=Q\,,\ \ \ \ \ \ \, \dot\rho_{\CC}=-{Q}\,.
\end{equation}
The form of the source function $Q$ depends on the assumptions made on the dynamical vacuum model (DVM).
Let us consider three model possibilities, one is  the ``running vacuum model''  (RVM) -- see \,\cite{JSPReview2013,SolGomRev2015,AdriaPhD2017} and references therein --  and the other two are called the $Q_{dm}$ and $Q_{\CC}$ models, whose names bear relation to the structure of the interaction source. For convenience they can also be labelled  I, II and III:
\begin{eqnarray}
&&{\rm Model\ I\ \ }({\rm RVM}):\ Q=\nu\,H(3\rho_{m}+4\rho_r)\label{eq:QforModelRVM} \label{eq:QforModelRVM}\\
&&{\rm Model\ II\ \ }(Q_{dm}):\ Q_{dm}=3\nu_{dm}H\rho_{dm} \label{eq:QforModelQdm}\\
&&{\rm Model\ III\ \ }(Q_{\CC}):\ Q_{\CC}=3\nu_{\CC}H\rho_{\CC}\,.\label{eq:QforModelQL}
\end{eqnarray}
Each model is characterized by a (dimensionless) coupling parameter $\nu_i=\nu,\nu_{dm},\nu_{\CC}$ in the interaction source, to be fitted to the observational data.  In the RVM case, the source function $Q$ in (\ref{eq:QforModelRVM}) can be derived from the following vacuum energy density:
\begin{equation}\label{eq:RVMvacuumdadensity}
\rho_\CC(H) = \frac{3}{8\pi{G}}\left(c_{0} + \nu{H^2}\right)\,,
\end{equation}
which can be motivated in QFT, see \cite{JSPReview2013}\,\footnote{The structure of the vacuum energy density in the running vacuum model can also be motivated from the effective behavior of the power-law solutions of Brans-Dicke gravity, see \cite{BD2018}.}. The other two models do not possess an analogous property and therefore are more {\it ad hoc}.
The additive constant $c_0$ in (\ref{eq:RVMvacuumdadensity}) is fixed by the boundary condition $\rho_\Lambda(H=H_0)=\rho^0_\Lambda$, where $H_0$ is the current Hubble rate and $\rho^0_\Lambda$ is the present measured value for $\rho_\Lambda$. The dimensionless parameter $\nu$ encodes the dynamics of the vacuum and can be related with the $\beta$-function of the $\rL$-running\,\cite{JSPReview2013}. Therefore we naturally expect $|\nu|<<1$ in QFT. Here we will treat $\nu$ as a free parameter of the RVM and, as indicated, it will be fitted to the data.

\begin{figure}[t]
\centering
\includegraphics[height=8.6cm]{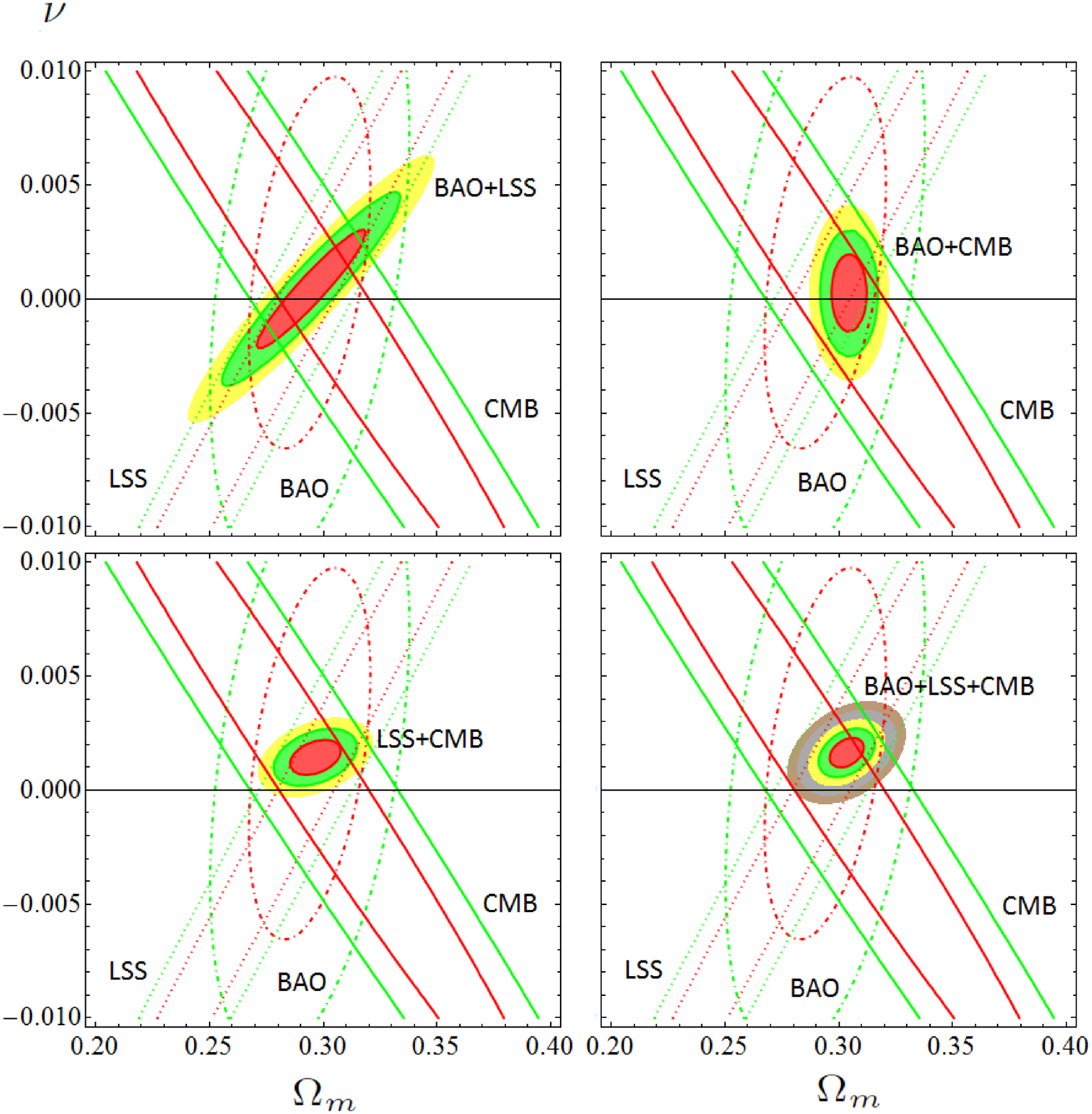}
\caption{Contour lines for the RVM  by considering the effect of only  BAO, LSS and CMB in all the possible pairings. Same data as in Fig.\,1. Remarkably the simultaneous combination of these three observables is able to capture signs of running vacuum at a confidence level of  $\gtrsim 3\sigma$ .}
\end{figure}
In Fig.\,1 we display the fitting results on the $(\Omega_m,\nu_i)$-planes for the three models I, II and III (corresponding to $\nu_i=\nu,\nu_{dm},\nu_{\CC}$), where $\Omega_m$ is the usual matter parameter at present. In Fig.\,2 we can better appraise the impact of the various data sources on these contour plots in the specific case of the RVM, see \cite{MNRAS-EPL2018} for details. The DDE option ($\nu\neq 0$, actually $\nu>0$) is clearly favored in the last plot when we combine the crucial triad of BAO+LSS+CMB data.

\section{Analytical solution of the dynamical vacuum models}

One can solve Models I, II and III  analytically in terms of the scale factor\,\cite{MNRAS-EPL2018}. The DM density in the case of the RVM  (Model I) as a function of the scale factor is
\begin{eqnarray}\label{eq:rhomRVMc}
\rho_{dm}^{(I)}= \rho_{dm}^0\,a^{-3(1-\nu)} + \rho_{b}^0\left(a^{-3(1-\nu)}
 - a^{-3}\right)- \frac{4\nu\rho^{0}_r}{1+3\nu}\left(a^{-4}-a^{-3(1-\nu)}\right)
\end{eqnarray}
and that of the vacuum energy density reads
\begin{eqnarray}\label{eq:rLRVMc}
\rho^{(I)}_\Lambda=\rLo + \frac{\nu}{1-\nu}\left[\rho^0_m\left(a^{-3(1-\nu)}-1\right)
 + \rho^{0}_{r}\left(\frac{(1-\nu)\,a^{-4} + 4\nu a^{-3(1-\nu)}}{1+3\nu}\,-1\right)\right].
\end{eqnarray}
The total matter density is just the sum of the conserved baryonic and radiation densities and the above DM density. As can be easily checked, for $\nu\to 0$ we recover the corresponding results for the $\CC$CDM, as it should.

For Model II:
\begin{eqnarray}\label{eq:rhoQdm}
\rho_{dm}^{(II)}(a) = \rho_{dm}^0\,a^{-3(1-\nu_{dm})}
\end{eqnarray}
and

\begin{eqnarray}\label{eq:rhoVQdm}
\rho_\CC^{(II)}(a) = \rLo + \frac{\nu_{dm}\,\rho_{dm}^0}{1-\nu_{dm}}\,\left(a^{-3(1-\nu_{dm})}-1\right)\,.
\end{eqnarray}
And for Model III:
\begin{eqnarray}\label{eq:rhoQL}
\rho_{dm}^{(III)}(a) =\rho_{dm}^0\,a^{-3} + \frac{\nu_\CC}{1-\nu_\CC}\rLo\left(a^{-3\nu_\Lambda}-a^{-3}\right)
\end{eqnarray}
and
\begin{eqnarray}\label{eq:rhoVQL}
\rho_\CC^{(III)}(a) =\rLo\,{a^{-3\nu_\CC}}\,.
\end{eqnarray}
Let us also briefly consider the simplest parametrization for DDE, namely  the XCDM\,\cite{XCDM}, in which the DE density as a function of the scale factor is simply given by $\rho_X(a)=\rho_{X0}\,a^{-3(1+w_0)}$, with $\rho_{X0}=\rho_{\CC 0}$. Here $w_0$ is the (constant) EoS parameter.
Last, but not least, we also study a scalar field model, denoted as $\phi$CDM. In this type of models the DE has a well-defined local Lagrangian description in terms of a scalar field. Taking $\phi$ dimensionless the energy density and pressure read
%%%
\begin{equation}
\rho_\phi = \frac{M^2_P}{16\pi}\left(\frac{\dot{\phi}^2}{2} + V(\phi) \right) \qquad p_\phi = \frac{M^2_P}{16\pi}\left(\frac{\dot{\phi}^2}{2} - V(\phi) \right),
\end{equation}
%%%
where $M_P$ is the Planck mass, which in natural units takes the value $M_P$ = $1/\sqrt{G}$ = 1.2211$\times 10^{19}$ GeV. Note that in our conventions $V(\phi)$ has dimension 2 in natural units. The scalar field satisfies the Klein-Gordon equation in the context of the FLRW metric $\ddot{\phi} + 3H\dot{\phi} + \partial{V(\phi)}/\partial\phi =0$. Here, as a particular case, we consider the original quintessence potential by Peebles and Ratra\,\cite{PeeblesRatra1988}:
\begin{equation}
V(\phi) = \frac{1}{2}\kappa{M^2_P}\phi^{-\alpha},
\end{equation}
%%%
where $\kappa$ and $\alpha$ are dimensionless. The parameter $\alpha$ should be positive and sufficiently small so that $V(\phi)$ can mimic approximately a CC slowly decreasing with time, such that it can eventually surface (near our time)  over the rapidly decaying matter density. In Fig. 3 we can see the corresponding contour lines for both the XCDM and $\phi$CDM, using SNIa+BAO+$H(z)$+LSS data (including the bispectrum component for the LSS+BAO data from \,\cite{GilMarin2016}) and compressed CMB Planck 2018 likelihood. The DDE option is consistently favored in both cases since $w_0>-1$ (for the XCDM) and $\alpha>0$ (for the $\phi$CDM)  at $\lesssim 3\sigma$ c.l. after marginalizing over $\Omega_m$\,\footnote{We should point out that a similar level of evidence is obtained using Planck 2015 data with full (i.e. non-compressed) CMB likelihood, see \cite{PDU2019} for details.}. Failure of including the bispectrum component (which seems to be particularly sensitive to the dynamics of the DE) may explain why other recent analyses on interacting vacuum and $\phi$CDM could not trace signs of DDE, see e.g. \cite{CostaLi,ParkRatraOoba}.
\newline
\newline

%%%%%%%%%%%%%%%%%%%%%%%%%%%%%%%%%%%%%%%%%%%%%%%%%%%%%%%%%%%%
\begin{figure}[t]
\centering
\includegraphics[angle=0,height=0.55\linewidth]{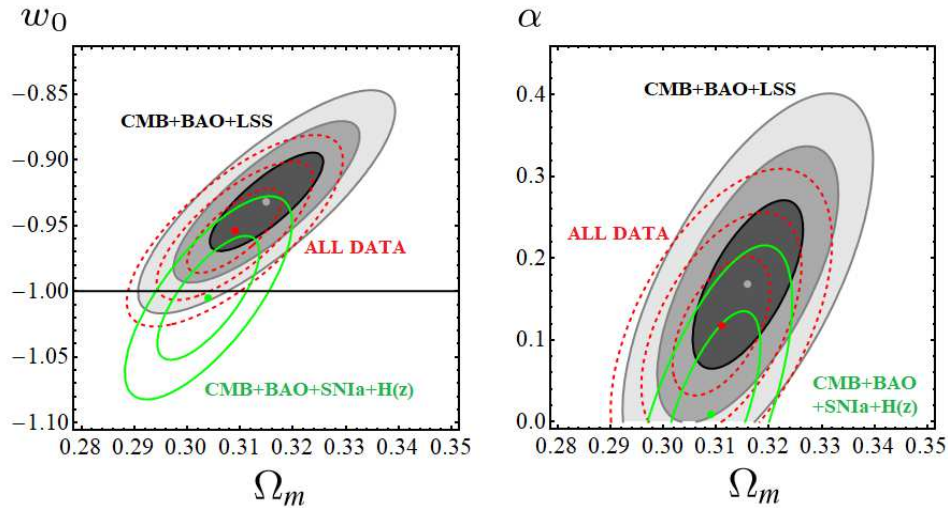}%\includegraphics[angle=0,height=0.4\linewidth]{CLPhiCDM.eps}
\caption{\label{fig:XCDMEvolution}%
\scriptsize Contour plots for the XCDM parametrization (left) and $\phi$CDM (right), using all sources of data (including LSS with bispectrum) as compared to  using all data except LSS. Here we use Planck 2018 data with compressed CMB likelihood. When LSS (with bispectrum)  is included, DDE  appears once more favored at  $\sim 3\sigma$ c.l. See \cite{PDU2019} for details.
%\vspace{0.3cm}
}
\end{figure}

\section{Dealing with the tensions through vacuum dynamics}

As indicated in the introduction, one of the persisting tensions with the $\CC$CDM is the discrepancy between the Planck value of the Hubble parameter $H_0$ obtained from the CMB anisotropies, and the local HST measurement (based on distance ladder estimates from Cepheids). The latter reads $H_0 = 73.24\pm 1.74$\, km/s/Mpc (referred to as R16) and $H_0 = 73.48\pm 1.66$\, km/s/Mpc (R18)\,\cite{RiessH0201618}. In contrast, the CMB value is $H_0 = 67.51\pm 0.64$ km/s/Mpc, as extracted from Planck 2015 TT,TE,EE+lowP+lensing data, or
$H_0 = 66.93 \pm 0.62$ km/s/Mpc, based on Planck 2015 TT,TE,EE+SIMlow data\,\cite{Planck2015}. In both cases there is a tension above $3\sigma$ c.l. with respect to the local measurement. This controversial situation has stimulated a lot of discussion in the literature. For instance, in\,\cite{Melchiorri2017b} the phantom DE option is exploited as a means to solve the tension.  In contrast, in\,\cite{PLB2017} it is proposed that vacuum and quasivacuum dynamics support the Planck measurement of $H_0$ against the local measurement with no need of phantom behavior.  A similar conclusion is reached with the XCDM and $\phi$CDM models studied here\,\cite{PDU2019}. In addition, a variety of model-independent approaches such as the Inverse Distance  Ladder method or the Multi-Task Gaussian Process\,\cite{H0Feeney} as well as other techniques (e.g. the Weighted Polynomial Regression\,\cite{LucaAdria}), do also support a lower value of  $H_0$ in the ballpark of Planck CMB measurements and hence in full consistency  with DDE.

%%%%%%%%%%%%%%%%%%%%%%%%%%%%%%%%%%%%%%%%%%%%%%%%%%%%%%%%%%%%%%%%%%%%%%%%%%%%%%%

\begin{figure}[t]
\begin{center}
\label{contours}
\includegraphics[width=4.5in]{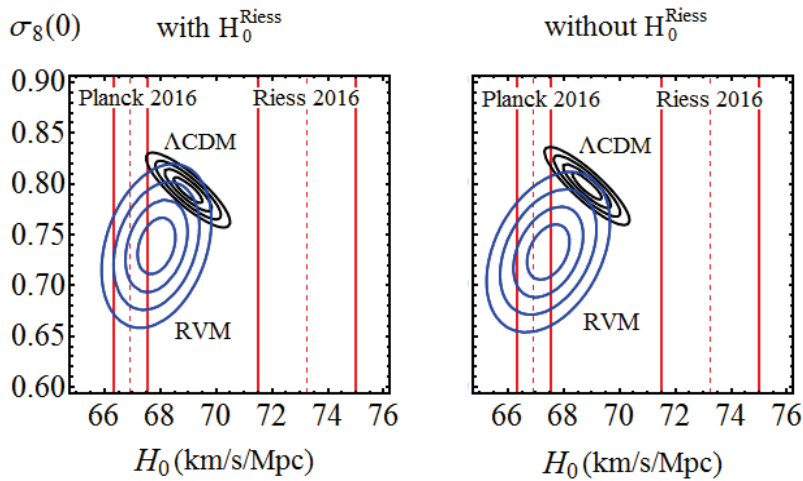}
\caption{\scriptsize Contour lines for the $\CC$CDM (black) and RVM (blue) up to $4\sigma$ in the $(H_0,\sigma_8(0))$-plane\,\cite{PLB2017}. On the left plot  we have included the R16 point\,\cite{RiessH0201618} on $H_0$ in our fit, whereas on the right plot it was not included. In both cases the favored range for $H_0$ is the Planck range. At the same time the optimal (lower) $\sigma_8(0)$ values are attained by the RVM, see Eq.\,(3), whereas the $\CC$CDM yields too large values.}
\end{center}
\end{figure}

%%%%%%%%%%%%%%%%%%%%%%%%%%%%%%%%%%%%%%%%%%%%%%%%%%%%%%%%%%%%%%%%%%%%%%%%%%%%%%%
The $H_0$-tension is closely connected with another type of problem, which is the role played by the LSS data, namely the data on $f(z)\sigma_8(z)$.   There seems to be no way at present to correctly account for both the CMB and the LSS data within the  $\CC$CDM. This new pitfall is at the root of the so-called  $\sigma_8$-tension, one of the most intriguing phenomenological  problems of the $\CC$CDM\,\cite{Macaulay2013,BasilakosNesseris2017}.  The problem has been dealt with in the literature from a variety of points of view, see e.g.\cite{Anand2017,AnFengWang2017}. Here we will focus once more on the possibility of DDE\,\cite{PLB2017,MNRAS2018a,MNRAS-EPL2018,PDU2019}.

In the presence of interacting vacuum  the perturbations equations of the $\CC$CDM must be modified appropriately\,\cite{MNRAS2018a,MNRAS-EPL2018}.
The detailed analysis of these matters performed in the mentioned references provides the basis for a possible solution to the $\sigma_8$-tension within the RVM.
In Fig.\,4, we present the likelihood contours in the $(H_0,\sigma_8(0))$-plane. In the left plot we show the situation when the local $H_0$ value R16 is included in the fit, whereas in the right plot we show the case when that local value is {\it not} included. The plots for the more updated R18 value would be almost indistinguishable from the R16 ones. From Fig.\,4 we can see that in order to reach the R16 neighborhood  the contours should be extended beyond the $5\sigma$ c.l., which would lead to a too large value of $\sigma_8(0)$ for the RVM. One can also show that it would lead to a too low value of $\Omega_m$\,\cite{PLB2017}.

Notice that $H_0$ and $\sigma_8(0)$ are positively correlated in the RVM (i.e. $H_0$ increases/decreases when $\sigma_8(0)$ increases/decreases, respectively), whilst they are anticorrelated in the $\CC$CDM ($H_0$ increases when $\sigma_8(0)$ decreases, and vice versa). The simultaneous $H_0$ and $\sigma_8(0)$ tensions could only be resolved simultaneously if the contours are extended well beyond $5\sigma$ at the expense of values of $\Omega_m$ below $0.27$. However, this does not seem to be the most likely solution. Vacuum dynamics offers a better scenario. The opposite correlation feature with respect to the $\CC$CDM allows e.g. the RVM to improve the LSS fit in the region where both $H_0$ and $\sigma_8(0)$ are smaller than the respective $\CC$CDM values (cf. Fig. 4). This explains why the Planck range for $H_0$ is clearly preferred by the RVM, as it allows a much better description of the LSS data. One can also derive similar conclusions using the XCDM and $\phi$CDM, see\cite{PDU2019}, where again a positive correlation between $H_0$ and  $\sigma_8(0)$   is found and the LSS tension can then be relaxed in the range of Planck values of $H_0$.

\section{Conclusions}

We have analyzed the ability of dynamical dark energy (DDE) models to fit the overall cosmological observations as compared to the $\CC$CDM. We find that there is non-negligible evidence that these models are more favored. Use of BAO+LSS data including the matter bispectrum is important to reach this conclusion. Furthermore, DDE is capable of smoothing out the LSS tensions of the concordance model with the data, and they favor the (lower) Planck value of $H_0$ over the local one.

\section{Acknowledgments}
We are funded by projects  FPA2016-76005-C2-1-P (MINECO), 2017-SGR-929 (Generalitat de Catalunya) and MDM-2014-0369 (ICCUB). JdCP  also by a  FPI fellowship associated to FPA2016.

\end{document}